\documentclass[twocolumn,showpacs,showkeys,preprintnumbers,amsmath,amssymb]{revtex4-1}
\usepackage{graphicx}% Include figure files
\usepackage{dcolumn}% Align table columns on decimal point
\usepackage{bm}% bold math

\begin{document}

%\preprint{Submitted to "J.\ Geophys.\ Res."}  % APS/123-QED

\title[]{Ionospheric current system accompanied by auroral vortex streets}

\author{Yasutaka Hiraki}
 \email{yhiraki28@gmail.com}
% \author{Keisuke Hosokawa}
\affiliation{
University of Electro-Communications, 1-5-1 Chofugaoka, Chofu, Tokyo, 182-8585, Japan.\\
}

%\author{Kaori Sakaguchi}
%\affiliation{
%National Institute of Information and Communications Technology, Tokyo, Japan.\\
%}

\date{\today}% It is always \today, today, but any date may be explicitly specified

\begin{abstract}
High resolution optical measurements have revealed that a sudden brightening of aurora and its deformation from an arc-like to a vortex street structure appear just at the onset of substorm. The instability of Alfv$\acute{\rm e}$n waves reflected from the ionosphere has been studied by means of magnetohydrodynamic simulations in order to comprehend the formation of auroral vortex streets. Our previous work reported that an initially placed arc intensifies, splits, and deforms into a vortex street during a couple of minutes, and the prime key is an enhancement of the convection electric field. This study elaborated physics of the ionospheric horizontal currents related to the vortex street in the context of so-called Cowling polarization. One component is due to the perturbed electric field by Alfv$\acute{\rm e}$n waves, and the other is due to the perturbed electron density (or polarization) in the ionosphere. It was found that, when a vortex street develops, upward/downward pair currents in its leading/trailing sides balance with an westward polarized component of the Hall current; it generates an eastward perturbed component of the Pedersen current. It was also found that both the perturbed component of the Hall current and the polarized component of the Pedersen current point equatorward, penetrating between the pair currents. 
\end{abstract}

%\pacs{94.30.Aa, 94.30.cq, 94.30.Lr}% PACS, the Physics and Astronomy
                             % Classification Scheme.
\keywords{Ionospheric current -- Cowling polarization -- MI coupling -- Alfv$\acute{\rm e}$n wave}%Use showkeys class option if keyword
                              %display desired
\maketitle

\section{Introduction}\label{sec: 1}
The auroral vortex street structure with a scale of 30--70 km has been investigated as a key to resolve the onset mechanisms of large-scale magnetospheric deformation or auroral breakup called substorm [Donovan et al., 2007; Sakaguchi et al., 2009]. From observations of Geotail and THEMIS satellites, it has been known that, even without any external input, plasmas are heated up to produce strong flow perturbations at a radial distance of $<10$ $R_{\rm E}$ that corresponds to the auroral region [Zesta et al., 2000; Keiling et al., 2009; Lee et al., 2012]. It is an established theory that an enhancement of plasma pressure triggers some instabilities in the magnetosphere, and field-aligned currents flow into the ionosphere, causing formation of bright auroral vortices [e.g., Lui et al. 2008]. % Refs, 要確認. 
It has been suggested that such a coherent structure in aurora is involved in magnetohydrodynamic (MHD) waves, especially, Alfv$\acute{\rm e}$n waves, that develops along the field line [Samson et al., 1992; Erickson et al., 2000]. Three-dimensional MHD simulations including magnetosphere-ionosphere (M-I) coupling were vigorously performed to interpret the vortex formation in aurora [Jia and Streltsov, 2014; Hiraki, 2015a, b]. 

Jia and Streltsov [2014] treated propagation and feedback instability of Alfv$\acute{\rm e}$n waves initiated by arc-like ionospheric density perturbation. They suggested that various structures form in aurora, or field-aligned currents, through changes in a set of controlled parameters in the system. However, their results showed that the time scale of auroral intensification and horizontal propagation of vortices reaches 200 s, which is still longer than the observed scale ($<60$ s). Furthermore, the value of field-aligned currents was smaller by over one-order than the estimated values (tens of $\mu$A/m$^2$) in observations [e.g., Dubyagin et al., 2003]. The lower estimate could mainly originate from the initial condition (or stability of the perturbations); more unstable eigenmodes of Alfv$\acute{\rm e}$n waves may control the system. 

On the contrary, Hiraki [2015b] performed a linear analysis of the feedback instability for realistic sets of convection electric field and conductivity and clarified the features of Alfv$\acute{\rm e}$n eigenmodes with maximum growth rates. Their MHD simulations demonstrated that the auroral vortex street develops in a fast time scale of 30--40 s, with a horizontal scale of 20--40 km and a strong field-aligned current of 10--20 $\mu$A/m$^2$. Splitting and dimming (a decrease in upward currents) of auroral arcs appear through a bounce motion of Alfv$\acute{\rm e}$n waves. A vortex street forms just after that, leading to a rapid intensification of arcs (i.e., an increase in upward currents). The sequence of these phenomena is consistent with substorm-related processes in past observations: fading, bead structure, and auroral breakup [e.g., Mende et al., 2009]. 

When a field-aligned current flows out from the ionosphere (especially, E-layer), the cold electron flux increases to cause a density fluctuation, and a new electric field anti-parallel to the ambient field is produced. The current produced by this "charge-up" process is often called as polarization current, or the Cowling channel [Baumjohann, 1983; Haerendel, 2008; Amm et al., 2013; Yoshikawa et al., 2013]. Previous studies have suggested that auroral structuring is attributed to this polarization current, especially due to a divergence of the Hall current [Buchert and Budnik, 1997; Amm and Fujii, 2008]. Fujii et al.\ [2012] presented a new model for the behavior of auroral arcs. Considering an arc with a finite length, an east-west directed electric field can be produced by the Cowling polarization. They pointed out that the polarization electric field creates a fast southward $E\times B$ drift of arcs, which has been often seen at the onset of substorms. 

Apart from auroral structures, Jia and Streltsov [2014] and Hiraki [2015a, b] treated all components of the electric field perturbation, the density perturbation, and these product (nonlinear term) of the ionospheric Pedersen/Hall currents. These are generally defined as a full set of the Cowling polarization currents [Yoshikawa et al., 2013]. 

No complete understanding has not been presented for ionospheric horizontal currents accompanied by the auroral vortex street. Our previous study [Hiraki, 2015b] showed only the temporal variation in the Pedersen/Hall currents associated with growth of Alfv$\acute{\rm e}$n waves, but did not interpret physics of the current balance. The first purpose of this study is to revisit the physics of vortex formation on the view point of horizontal currents. Different from the quantities of Alfv$\acute{\rm e}$n waves, the advantage of the Pedersen/Hall currents is to be measured directly by ground magnetometer and ionospheric convective flow observations. The analysis of growing modes expected that the vortex street motion points nearly parallel to the ionospheric horizontal (Pedersen + Hall) currents. This theoretical speculation could be compared with precise optical observations that can determine the direction of wavefronts. On the other hand, a statistical analysis by Fujii et al.\ [2012] signified that the speed of auroral arcs is independent of the ambient convection speed. The second purpose of this study with exact numerical simulations is to verify their scenario of auroral intensification by the Cowling effect, although the model setup is somewhat different.

\section{Model Description}\label{sec: 2}
The same equations as Hiraki [2015b] are used to describe the motion of shear Alfv$\acute{\rm e}$n waves involved in auroral structuring. Shear Alfv$\acute{\rm e}$n waves penetrate into the ionosphere and are destabilized due to an enhancement of plasma convection. The equations for ionospheric plasma are also the same as those of Hiraki [2015b]. The outline of the coordinate system and equations of our scope is explained as follows. 

We take into account the locally orthogonal coordinates (${\bm x}(s)$, ${\bm y}(s)$, ${\bm s}$) in the dipole magnetic field; the relation of each unit vector is ${\bm e}_x \times {\bm e}_y = {\bm e}_s$. The field line position $s$ is defined as $s = 0$ at the ionosphere and $s = l$ at the magnetic equator. We consider a latitude of $70^\circ$ in the southern hemisphere, with the dipole $L$ value of $\approx 8.5$ and $l \approx 7\times10^4$ km. The coordinate ${\bm x}$ points southward (poleward) and ${\bm y}$ points eastward at $s = 0$. We set a local flux tube, e.g., a square of ($l_\perp \times l_\perp$) at $s = 0$ and a rectangle of ($\approx 3300$ km $\times$ $\approx 1700$ km) at $s = l$ using $l_\perp \approx 70$ km and dipole metrics; see Hiraki and Watanabe [2011] for details. 

The dipole magnetic field is written as ${\bm B}_0$. The system has a convective electric field ${\bm E}_0$ ($\perp {\bm B}_0$) that is applied poleward ($\parallel {\bm x}$) and is uniform in every $x$-$y$ planes. With perturbed electric and magnetic fields of ${\bm E}_1 = - B_0 {\bm \nabla}_\perp \phi$ and ${\bm B}_1 = {\bm \nabla}_\perp \psi \times {\bm B}_0$, the reduced MHD equations for shear Alfv$\acute{\rm e}$n waves are expressed as 
\begin{eqnarray}
 \displaystyle && \partial_t \omega + {\bm v}_\perp \cdot {\bm \nabla}_\perp \omega = v_{\rm A}^2 \nabla_\parallel j_\parallel  \\
 && \partial_t \psi + {\bm v}_0 \cdot {\bm \nabla}_\perp \psi + \frac{1}{B_0} \nabla_\parallel B_0 \phi = -\eta j_\parallel. 
\end{eqnarray}
The domain of definition is $0 < s \le l$. Here, $\omega = \nabla_\perp^2 \phi$ stands for vorticity, $j_\parallel = - \nabla_\perp^2 \psi$ field-aligned current, ${\bm v}_\perp = {\bm v}_0 + {\bm v}_1$, ${\bm v}_{0, 1} = {\bm E}_{0, 1} \times {\bm B}_0 / B_0^2$ the convective and perturbed drift speeds, $v_{\rm A}$ the Alfv$\acute{\rm e}$n velocity, $\eta$ resistivity, and $\nabla_\parallel = \partial_s + {\bm b}_0 \cdot {\bm \nabla}_\perp \times {\bm \nabla}_\perp \psi$. The $v_0$ is set so that $E_0$ satisfies the equipotential condition along the field line. Suppose that upward $j_\parallel$ represents the shapes of aurora in this study. 

By integrating the continuity equations of ions and electrons over the current dynamo layer (height of 100--150 km), 
equations that represent the ionospheric plasma motion at $s = 0$ are 
\begin{eqnarray}
 \displaystyle && \partial_t n_{\rm e} + {\bm v}_\perp \cdot {\bm \nabla}_\perp n_{\rm e} = j_\parallel - R n_{\rm e} \\
 && {\bm \nabla}_\perp \cdot (n_{\rm e} \mu_{\rm P} {\bm E}) - {\bm v}_\perp \cdot {\bm \nabla}_\perp n_{\rm e} = D \nabla_\perp^2 n_{\rm e} - j_\parallel. 
\end{eqnarray}
Equation (3) is the continuity equation of electron itself, and its velocity is assumed as ${\bm v}_{\rm e} = {\bm v}_\perp$; electrons yield the Hall drift. The electron density is partitioned into $n_{\rm e}({\bm x}_\perp,t) = n_0 + n_1({\bm x}_\perp,t)$. We assume that the field-aligned current is carried by thermal electrons and is treated equal to $j_\parallel$ shown in Eqs.\ (1) and (2). The linearized recombination term is written as $R n_{\rm e}$. By imposing charge quasi-neutrality, equation (4) is given when eq.\ (3) is subtracted from the continuity equation of ions. The ion velocity is written as ${\bm v}_{\rm i} = \mu_{\rm P} {\bm E}$; ions yield the Pedersen drift. The Hall mobility $\mu_{\rm H}$ appearing in Eqs.\ (3) and (4) was normalized to be unity, and $D$ is the diffusion coefficient. Ionospheric density waves governed by Eqs.\ (3) and (4) couple to shear Alfv$\acute{\rm e}$n waves. 

The numerical schemes solving Eqs.\ (1)--(4) and the values of parameters were the same as Hiraki [2015b]. The fourth-order central difference method in space and the fourth-order Runge-Kutta-Gill method in time were used for Eqs.\ (1)--(3). The multigrid-BiCGSTAB method was used for Eq.\ (4). The number of grids was (256, 256, 128) for the ${\bm x}(s)$, ${\bm y}(s)$, and ${\bm s}$ directions, respectively; thus, $\Delta x \approx 0.27$ km since $l_\perp = 70$ km at $s = 0$. The time resolution was changed in accord with the Courant condition. The numerical viscosity and resistivity equaled $10^{-7}/B_0$. We set a periodic boundary in the ${\bm x}$ and ${\bm y}$ directions. The potential was taken from Eq.\ (4) at the boundary of $s = 0$, while the anti-symmetric boundary condition for magnetic potential $\psi = 0$ (or $j_\parallel = 0$) was put at $s = l$. 

The Alfv$\acute{\rm e}$n velocity $v_{\rm A}$ is set to be constant as $\approx 1.5 \times 10^3$ km/s along the field line. The Alfv$\acute{\rm e}$n transit time from $s = 0$ to $s = l$ is thus $\tau_{\rm A} = \int_{0}^{l} 1/v_{\rm A}(s) {\rm d}s \approx 47$ s. The values of parameters at $s = 0$ are shown below. In this study, we consider a situation where a poleward convection electric field of $E_0 = 60$ mV/m is applied. As the magnetic field is $B_0 \approx 5.7 \times 10^{-5}$ T, the convection speed is $v_0 = E_0 / B_0 \approx 1.1$ km/s. The ambient density is $n_0\approx 3.8 \times 10^4$ cm$^{-3}$, $\mu_{\rm P}$/$\mu_{\rm H} = 0.5$, $\Sigma_{\rm P}/\Sigma_{\rm A} = 5$, $D = 4 \times 10^5$ m$^2$/s, and $R = 2 \times 10^{-3}$ s$^{-1}$. 

Setting the initial condition, field variables are partitioned into $(\phi, \psi, n_{\rm e}) = (\phi_{\rm a}, \psi_{\rm a}, n_{\rm ea}) + (\tilde{\phi}, \tilde{\psi}, \tilde{n}_{\rm e})$. The arc potential is yielded as the fundamental wave form of $\phi_{\rm a}(s)\propto \frac{1}{B_0} \sin (\frac{\pi}{2l} s)$ along the field line, while $\psi_{\rm a}(s) = n_{\rm ea} = 0$ for simplicity. The potential $\phi_{\rm a}$ is assumed to be Gaussian-like with a scale of 10 km and a peak value of 20 mV/m and satisfies the periodic boundary condition. In these settings, an auroral arc with $j_\parallel > 0$ quickly appears at the center of ${\bm x}$ coordinate. The perturbed variables $(\tilde{\phi}(s), \tilde{\psi}(s), \tilde{n}_{\rm e})$ were represented by eigenfunctions obtained from a linear set of Eqs.\ (1)--(4) [Hiraki and Watanabe, 2011; 2012]. The eigenfunctions with a wave number of ${\bm k}_\perp = (k_x, k_y) = (1, 2)$ provide the maximum growth rate for the above parameter setting (especially, $E_0 = 60$ mV/m); see Hiraki [2015b] for details. Here, wave number is normalized by $2 \pi / l_\perp$. The unstable initial perturbation has an amplitude of $10^{-4} |\phi_{\rm a}|$. 
% アーク周りに半時計回りの対流を想定

\begin{figure}[t]
\includegraphics[width=1.0\columnwidth, bb=0 0 360 252, clip]{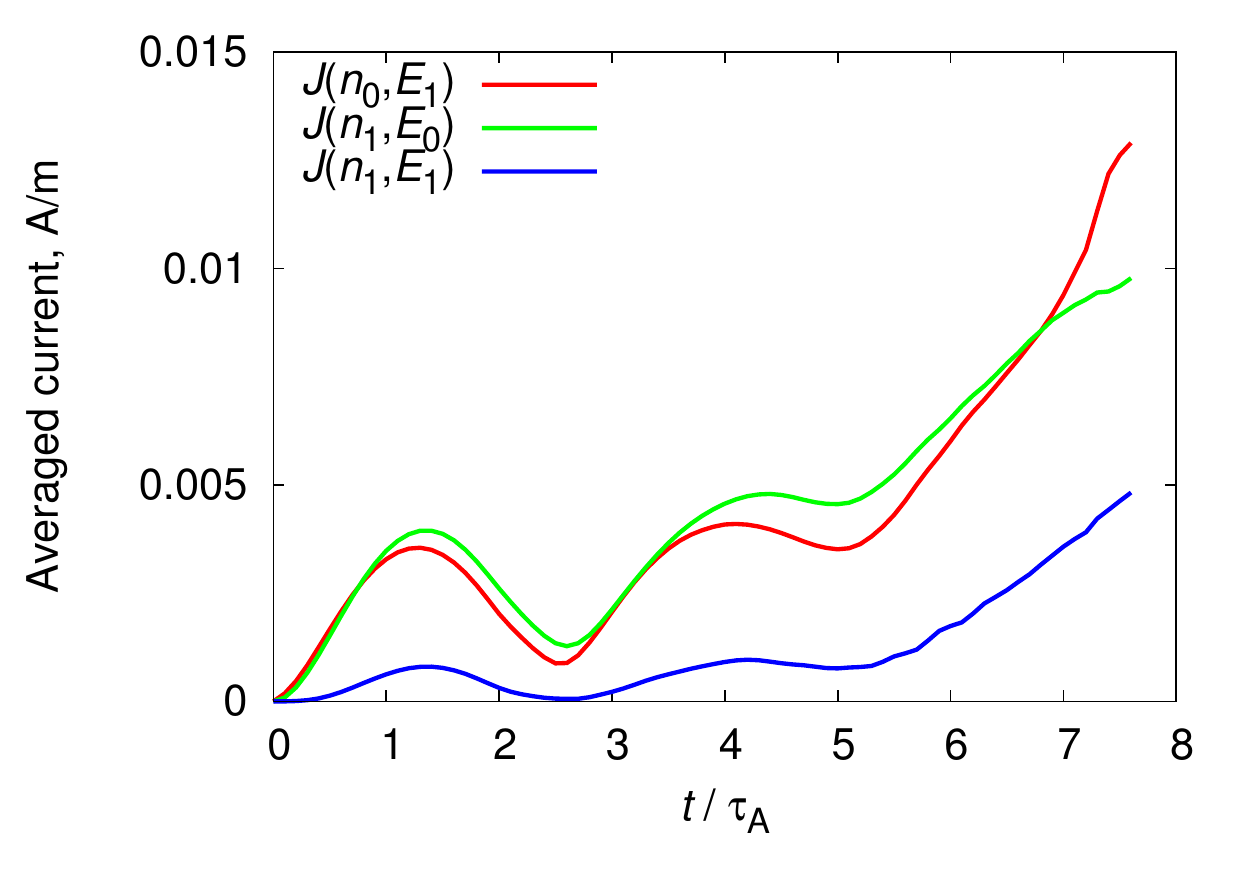}
\caption{Temporal variation in the total (Pedersen + Hall) current averaged in the ionosphere: $J(n_0, E_1)$, $J(n_1, E_0)$, and $J(n_1, E_1)$ stand for the perturbed, polarized, and nonlinear currents, respectively, and $\tau_{\rm A}$ is the Alfv$\acute{\rm e}$n transit time (see text).}
\end{figure}

\begin{figure*}[t]
\includegraphics[width=2.0\columnwidth, bb=130 40 1170 630, clip]{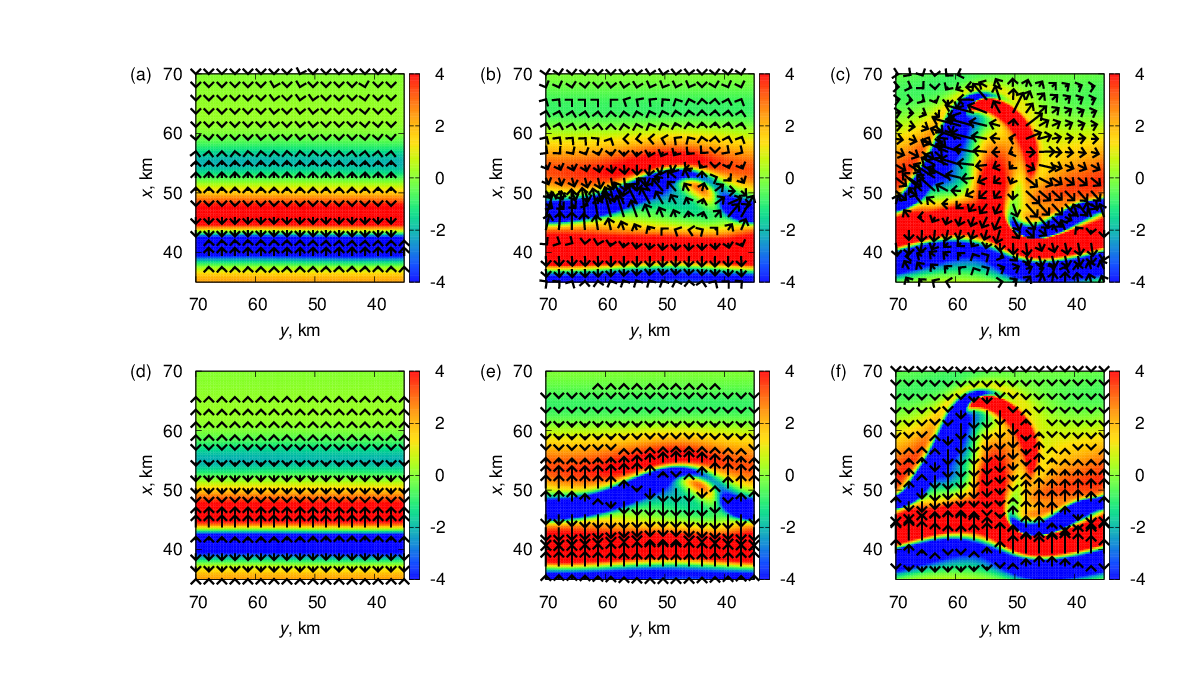}
\caption{Snap shots of spatial pattern of the field-aligned current (color contour, $\mu$A/m$^2$) and the Pedersen currents (vectors) at the ionosphere. Shown are the perturbed (a--c) and polarized (d--f) components at $t/\tau_{\rm A} = 5.5$, 7, and 7.6, respectively. For each period, $j_{\rm P, \max} \approx 0.067$, 0.091, and 0.14 A/m.}
\end{figure*}

\begin{figure*}[t]
\includegraphics[width=2.0\columnwidth, bb=130 40 1170 630, clip]{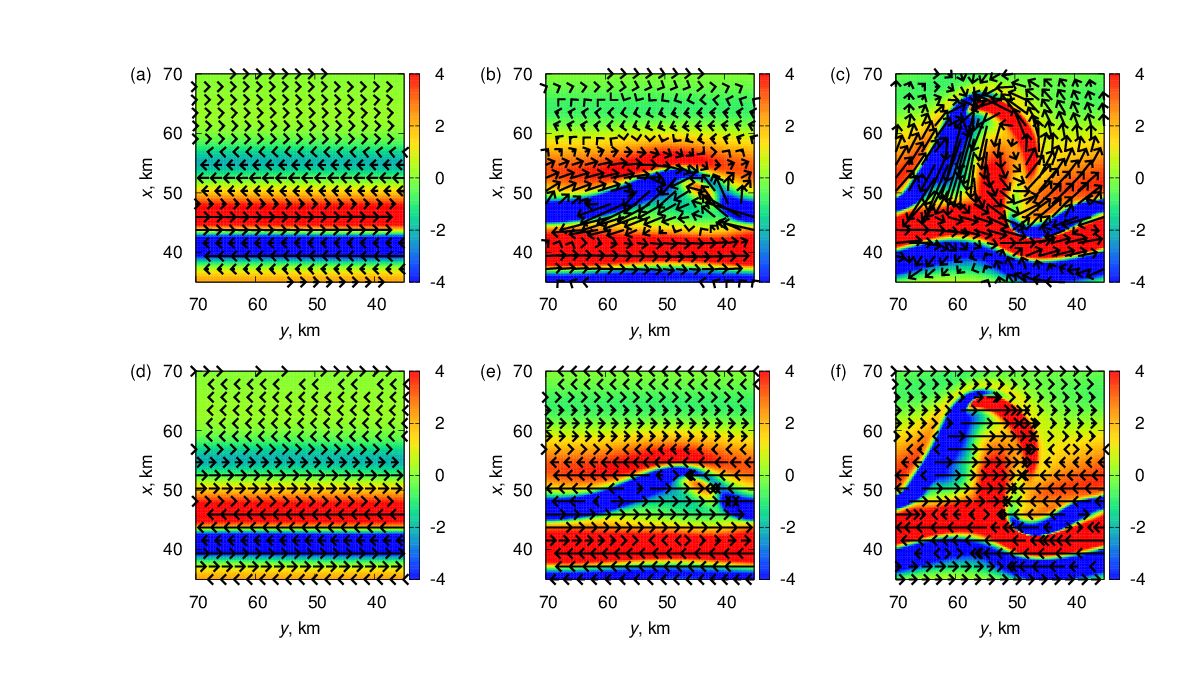}
\caption{Same as Fig.\ 2, but for the Hall current. For each period, $j_{\rm H, \max} \approx 0.13$, 0.18, and 0.28 A/m.}
\end{figure*}

\begin{figure*}[t]
\includegraphics[scale=0.55, bb=0 10 700 340, clip]{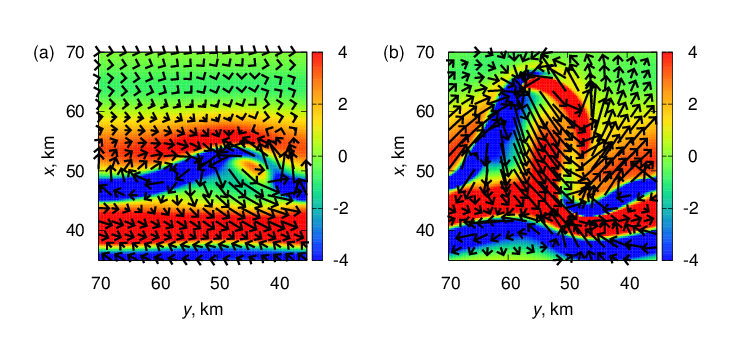}
\caption{Same as Fig.\ 2, but for the total current $j_{\rm P}$ + $j_{\rm H}$ at $t/\tau_{\rm A} = 7$ (a) and 7.6 (b). For each period, $J_{\max} \approx 0.20$ and 0.31 A/m.}
\end{figure*}

\section{Results}\label{sec: 3}
We performed a three-dimensional MHD simulation in the nominal case of parameters shown in Sec.\ \ref{sec: 2}. During a bounce motion between the ionosphere $s = 0$ and the magnetic equator $s = l$, the Alfv$\acute{\rm e}$n wave interacts with ionospheric density waves to increase their amplitudes by the feedback instability. Evolutions of vorticity and field-aligned current in this situation are referred to Hiraki [2015b]. This study elaborated the characteristics of ionospheric horizontal currents ${\bm j}_{\rm P, H}$. The velocities of ions and electrons, respectively, yield the Pedersen current ${\bm j}_{\rm P}$ and the Hall current ${\bm j}_{\rm H}$ as stated at Eqs.\ (3) and (4). Using the ordering of $n_{\rm e}$ and ${\bm E}$ in Sec.\ \ref{sec: 2}, these currents are expanded into 
\begin{eqnarray}
 \displaystyle {\bm j}_{\rm P} = n_{\rm e} {\bm v}_{\rm i} &=& (n_0 + n_1) \mu_{\rm P} ({\bm E}_0 + {\bm E}_1) \nonumber \\
 &=& \mu_{\rm P} (n_0 {\bm E}_0 + n_0 {\bm E}_1 + n_1 {\bm E}_0 + n_1 {\bm E}_1), \\
 {\bm j}_{\rm H} = -n_{\rm e} {\bm v}_{\rm e} &=& (n_0 + n_1) {\bm b}_0 \times ({\bm E}_0 + {\bm E}_1) \nonumber \\
 &=& {\bm b}_0 \times (n_0 {\bm E}_0 + n_0 {\bm E}_1 + n_1 {\bm E}_0 + n_1 {\bm E}_1). 
\end{eqnarray}
Here, note that the magnetic field $B_0$ and the Hall mobility $\mu_{\rm H}$ were normalized to be unity. The first term of Eqs.\ (5) and (6) stands for the ambient component of current, but does not concern formation in spatial (vortex) structures since it was set constant. The second term is the perturbed field component, and the third term is the polarized component due to the perturbed density. The fourth term is the nonlinear term but does not have any contribution to vortex formation in our case (see Fig.\ 1 and discussions below). 

Expressing the sum of horizontal currents as $J(n_{\rm i}, E_{\rm j}) = |{\bm j}_{\rm P} + {\bm j}_{\rm H}|(n_{\rm i}, E_{\rm j})$, figure 1 shows the temporal variation in root-mean-square values (A/m) of the 2nd term $J(n_0, E_1)$, the 3rd term $J(n_1, E_0)$, and the 4th term $J(n_1, E_1)$ of Eqs.\ (5) and (6). All terms have local maxima at $t/\tau_{\rm A} \approx 1.2$ and 4.1. It implies that the Alfv$\acute{\rm e}$n wave returns to the ionosphere, through a bounce motion, to amplify the electric field therein. After $t/\tau_{\rm A} \approx 5$, the current amplitudes rapidly grow in a time scale of $2 \tau_{\rm A} \approx 94$ s. Although formation of the vortex street at $t/\tau_{\rm A} = 6$--7 is clearly found from Figs.\ 2 and 3 (shown later), we can also infer the onset timing of vortex street from the average current $J(t)$. The perturbed field component $J(n_0, E_1)$ exceeds the ambient field (polarized) component $J(n_1, E_0)$ at $t/\tau_{\rm A} \approx 6.8$. It physically means that an eastward flow perturbation ${\bm E}_1 \times {\bm B}_0$ grows to compete with the ambient westward flow ${\bm E}_0 \times {\bm B}_0$ at the poleward side of an arc, and a vortex street forms due to the shear. The nonlinear term $J(n_1, E_1)$ also increases to be a $50 \%$ level of these two components. 

Figure 2 shows the distribution of the perturbed and polarized components, ${\bm j}_{\rm P}(n_0, {\bm E}_1)$ (a--c) and ${\bm j}_{\rm P}(n_1, {\bm E}_0)$ (d--f), of the Pedersen current; (a, d), (b, e), and (c, f) are plots at $t/\tau_{\rm A} = 5.5$, 7, and 7.6, respectively. Color contours show the field-aligned current $j_\parallel$ in units of $\mu$A/m$^2$ at $s = 0$. Note that, hereafter, plots are shown in the moving frame of ${\bm v}_0$ ($\parallel -y$, or westward). From $t/\tau_{\rm A} = 0$ to 5.5, a region of upward current (an auroral arc) imposed at $x = 35$ km gets on the Pedersen drift parallel to the southward electric field. The polarized component ${\bm j}_{\rm P}(n_1, {\bm E}_0)$ is diverged at a downward $j_\parallel$ while is converged at an upward $j_\parallel$, which means that so-called the Bostr$\ddot{\rm o}$m type pair current forms (see Fig.\ 2(d)) [Bostr$\ddot{\rm o}$m, 1964]. Since the local maximum of $|{\bm E}_1|$ is placed at the boundaries of $j_\parallel$, the pattern of the perturbed component ${\bm j}_{\rm P}(n_0, {\bm E}_1)$ is in anti-phase of ${\bm j}_{\rm P}(n_1, {\bm E}_0)$; it is converged at $j_\parallel < 0$ and is diverged at $j_\parallel>0$. A vortex forms at ($x$, $y$) = (50 km, 45 km) at $t/\tau_{\rm A} = 7$ to expand poleward until $t/\tau_{\rm A} = 7.6$. The above relationships in currents are still satisfied in the latter time, except regions of vortices. Vortices propagate westward (in the rest frame), and an upward $j_\parallel$ is produced at the leading side. A characteristic of the Pedersen current inside the vortex is that a divergence of ${\bm j}_{\rm P}(n_0, {\bm E}_1)$ exists at the leading side, the left part of which points to the trailing side (a region of $j_\parallel < 0$); we will discuss on the current balance in Sec.\ \ref{sec: 4}. 

Figure 3 shows the distribution of the perturbed and polarized components, ${\bm j}_{\rm H}(n_0, {\bm E}_1)$ (a--c) and ${\bm j}_{\rm H}(n_1, {\bm E}_0)$ (d--f), of the Hall current; times are the same as those in Fig.\ 2. Let us first see the polarized component ${\bm j}_{\rm H}(n_1, {\bm E}_0)$. Since local maximum and minimum in vorticity (or electric potential) form, counter-clockwise and clockwise patterns in ${\bm j}_{\rm H}(n_1, {\bm E}_0)$ are produced at regions of $j_\parallel < 0$ and $>0$, respectively; see Fig.\ 3(d). On the other hand, the perturbed component ${\bm j}_{\rm H}(n_0, {\bm E}_1)$ exhibits an anti-phase behavior; clockwise/counter-clockwise flows form at regions of $j_\parallel < 0$/$> 0$. Similar to the Pedersen current, the relation in currents are satisfied at $t/\tau_{\rm A} = 7$ and 7.6, except regions of vortices. A characteristic of the Hall current inside the vortex is that the polarized current flows from the trailing side ($j_\parallel < 0$) to the leading side ($j_\parallel > 0$), while the perturbed current exhibits clockwise/counter-clockwise patterns at each side. 

Instead of the nonlinear term ${\bm j}_{\rm P,H}(n_1, {\bm E}_1)$ almost subordinate to the polarities of $n_1$ and $E_1$, this paper concentrates on the behavior of the total current $\Sigma_{\rm (i, j) \ne (0, 0)} \{ {\bm j}_{\rm P}(n_{\rm i}, {\bm E}_{\rm j}) + {\bm j}_{\rm H}(n_{\rm i}, {\bm E}_{\rm j}) \}$, including ${\bm j}_{\rm P,H}(n_1, {\bm E}_1)$, that attracts much attention from the view point of observations. Figure 4 shows the 2D plots of the total current at $t/\tau_{\rm A} = 7$ and 7.6. At $t/\tau_{\rm A} = 7$, clockwise/counter-clockwise vortex currents form at regions of $j_\parallel < 0$ and a newly produced $j_\parallel > 0$, respectively (see $x \approx 50$ km). This is roughly interpreted to be a combination of the perturbed components, ${\bm j}_{\rm P}(n_0, {\bm E}_1)$ and ${\bm j}_{\rm H}(n_0, {\bm E}_1)$, because they exceed the polarized components at this period. On the other hand, a combination of the polarized components, ${\bm j}_{\rm P}(n_1, {\bm E}_0)$ and ${\bm j}_{\rm H}(n_1, {\bm E}_0)$, dominates at the arc region of upward current at $x\approx 40$ km. The separation of the perturbed and polarized components breaks at $t/\tau_{\rm A} = 7.6$. An equatorward current penetrating at the center of vortex ($y \approx 57$ km) is interpreted to be a composition of ${\bm j}_{\rm H}(n_0, {\bm E}_1)$ and ${\bm j}_{\rm H}(n_1, {\bm E}_0)$. A contribution of ${\bm j}_{\rm P}(n_0, {\bm E}_1)$ appears outside the vortex. Convergence/divergence patterns by a composite of the perturbed and polarized components emerge even at the arc region ($x\approx 42$ km). 
% それぞれ赤道/極に張り出した領域で

\section{Discussion}\label{sec: 4}
Let us discuss on the formation of Cowling polarization currents accompanied by calculated vortex streets in Sec.\ \ref{sec: 3} and further make a comparison with previous studies. In the following analysis, we removed the nonlinear component given by the 4th terms in Eqs. (5) and (6). 

\subsection{Interpretation of the current system in auroral vortex streets}
We assumed that coordinates ${\bm x}$ and ${\bm y}$ point southward and eastward, respectively, in the ionosphere; remember that we consider the southern hemisphere. The convection electric field ${\bm E}_0$ is induced to the $x$ direction. The 1st terms in Eqs.\ (5) and (6) (constant) occupy a much global scale than that of auroral structures. We consider a current sheet elongated along the $y$ direction. Before an auroral vortex street forms, we can assume $\partial_y = 0$ and thus $E_{1y} = - \partial_y \phi = 0$. In this condition and without the diffusion term in Eq.\ (4), we obtain the equation as 
\begin{equation}
 \displaystyle \mu_{\rm P} \partial_x (n_0 E_{1x} + n_1 E_0) = - j_\parallel, 
\end{equation}
which includes only the Pedersen current components. 

We return to Eq.\ (3) in order to interpret the results of the Pedersen current in Fig.\ 2. By neglecting the nonlinear advection and recombination terms, we find that the density initially changes to positive $n_1 > 0$ or negative $< 0$ at regions of upward/downward field-aligned currents ($j_\parallel >0$ or $<0$), respectively. i) The polarized current due to $n_1$, or the 2nd term in Eq.\ (7), primarily grows in that situation. The right-hand term being negative/positive in cases of upward/downward currents ($j_\parallel >0$ or $<0$) along with a positive constant $E_0$, the polarity of the left-hand 2nd term is $E_0 \partial_x n_1 <0$ or $> 0$. In the case of an upward current, a converged current system forms where $n_1 < 0$, or $j_{\rm P} < 0$, is produced at the poleward side of an arc, while $j_{\rm P} > 0$ at the equatorward side. A diverged current system forms for a downward current $j_\parallel < 0$. These correspond to the so-called Bostr$\ddot{\rm o}$m type current system [Bostr$\ddot{\rm o}$m, 1964; Haerendel, 2010]. ii) Suppose that the left-hand 2nd term induces the 1st term in Eq.\ (7), regions without $j_\parallel$ can be simply understood. Integrating in the periodic boundary condition, we obtain the relation $n_0 E_{1x} + n_1 E_0 = 0$. Remember that the 2nd term is negative/positive at the poleward/equatorward sides of an upward current, the 1st term is quickly found to be $j_{\rm P} > 0$ or $< 0$ as $n_0>0$. It means that the perturbed component flows anti-parallel to the polarized component. Figure 2 shows that these currents satisfy this relation at least until $t/\tau_{\rm A} = 7$. 

When the perturbed field grows to be $E_{1y} \ne 0$, the system turns to be more complex with the perturbed Hall current. Assuming a weak $y$ dependence as $\partial_y = \epsilon \partial_y$ and $E_{1y} = \epsilon E_{1y}$ along with $j_\parallel = j_1 + \epsilon j_2$, Eq.\ (4) is expanded to be Eq.\ (7) and 
\begin{equation}
 \displaystyle - n_0 {\bm b}_0 \cdot {\bm \nabla}_\perp \times {\bm E}_1 + E_0 \partial_y n_1 = - j_2. 
\end{equation}
The Hall current is produced by $j_2$, independent of the Pedersen current. Here, though the left-hand 1st term itself is zero, we keep it since ${\bm j}_{\rm H}(n_0, {\bm E}_1) \ne 0$. 

From the above analysis of Eq.\ (7), the current system at $t/\tau_{\rm A} = 5.5$ can be explained as a continuity of $j_\parallel$ and the ${\bm x}$ components of ${\bm j}_{\rm P}$. No new current $j_2$ is produced at this stage. As mentioned above at ii), $E_{1x} > 0$ or $< 0$ is produced at the poleward/equatorward sides of an upward current $j_1 >0$. The ${\bm y}$ component of the perturbed Hall current in the 1st term of Eq.\ (8) shows a counter-clockwise flow at regions of $j_1 > 0$, while a clockwise flow at $j_1 < 0$. These features are found in Fig.\ 3(a). If $E_{1y}$ is regarded relatively small in Eq.\ (8), we obtain $n_0 E_{1x} + n_1 E_0 = 0$, and thus the polarized current in Fig.\ 3(d) flows anti-parallel to the perturbed component. 

The variables $E_{1y}$, $\partial_y$, and $j_2$ develop at $t/\tau_{\rm A} = 7$. From the relation in Eq.\ (8), $E_0 \partial_y n_1 < 0$ meets at a patchy area of $j_2 > 0$ ($x \approx 50$ km), which means a converged current system: i.e., the current flows westward at the eastward side, and vice versa. The currents are diverged at $j_2 < 0$. Figure 3(e) suggests that the newly produced $j_2$ balances with the Hall current ${\bm j}_{\rm H}(n_1, {\bm E}_0)$. On the other hand, the perturbed component (Fig.\ 3(b)) develops into counter-clockwise/clockwise flows at regions of $j_2 > 0$ or $< 0$ as similar to patterns driven by $j_1$. The twin vortices, growing further at $t/\tau_{\rm A} =  7.6$ (Fig.\ 3(c)), are caused by a coupling of $j_2$ and the Hall current. The Hall current points to the $-y$ direction (clockwise) at the poleward side of a downward current $j_1 < 0$ at $x = 45$--49 km. A strong shear is produced by this current and the $y$-directing ambient Hall current (the 1st term in Eq.\ (6)). As a result, a counter-clockwise flow of ${\bm j}_{\rm H}$, or a localized new $j_2 > 0$, forms at ($x$, $y$) = (50 km, 45 km); see Figs.\ 3(b, e). 

As mentioned just above, we find that twin vortices of the perturbed component of the Hall current develop at $t/\tau_{\rm A} = 7.6$ associated with $j_\parallel$. However, the distribution of the Pedersen and Hall currents at $t/\tau_{\rm A} = 7.6$ is not explained simply by Eqs.\ (7) and (8). The perturbed and polarized components no longer point anti-parallel direction, but are almost normal to each other. It means generation of a coupling of the Pedersen and Hall currents. Let us consider the features inside of pair currents of upward/downward $j_\parallel$ (or twin vortices) at ($x$, $y$) = (50--60 km, 50--60 km). Since structures of $j_\parallel$ are stretched along the $x$ direction, we make a simple assumption as $\partial_x = \epsilon \partial_x$, and thus $E_{1x} = {\cal O}(\epsilon)$, in this region. With the relation $E_{1y}$, $j_2 = {\cal O}(1)$, Eq.\ (4) is expanded into 
\begin{eqnarray}
 \displaystyle && \partial_y (\mu_{\rm P} n_0 E_{1y} + n_1 E_0) = - j_2 \\
 && - n_0 {\bm b}_0 \cdot {\bm \nabla}_\perp \times {\bm E}_1 + \mu_{\rm P} E_0 \partial_x n_1 = 0. 
\end{eqnarray}
Here, the 1st term in Eq.\ (9) is the perturbed component of the Pedersen current, while the 2nd term is the polarized component of the Hall current. The 1st term in Eq.\ (10) is the perturbed component of the Hall current, while the 2nd term is the polarized component of the Pedersen current. 

The upward current $j_2$ at ($x$, $y$) = (50 km, 45 km) at $t/\tau_{\rm A} = 7$ promotes to a strong upward current as ${\cal O}(1)$ at ($x$, $y$) = (60 km, 50 km) at $t/\tau_{\rm A} = 7.6$. The left-hand 2nd term of Eq.\ (9) produces a convergence of the Hall current shown in Fig.\ 3(f). When $E_{1y}$ grows to be ${\cal O}(1)$, a divergence of the Pedersen current, anti-parallel to the Hall current, forms as shown in Fig.\ 2(c). These directions are reversed at the region of downward current at ($x$, $y$) = (60 km, 60 km). As for the balance in Eq.\ (10), $E_{1x} > 0$ dominates at the poleward side of the upward current from the stage of Eq.\ (7), and a strong flow in the $\pm x$ direction is produced through a growth of $E_{1y}$. By the left-hand 1st term of Eq.\ (10), twin vortices of the Hall current, or counter-clockwise/clockwise flows, form at regions of upward/downward currents in Fig.\ 3(c). We would further focus on the inside ($x$, $y$) = (55 km, 55 km) of these pair currents where there is a strong Hall current in the $-x$ direction. Assuiming $E_{1x}$ small in this region, Eq.\ (10) is turned to be $\partial_x ( -n_0 E_{1y} + \mu_{\rm P} n_1 E_0 ) = 0$. It is roughly interpreted that, due to polarization $n_1 < 0$ [Hiraki, 2015b], the polarized Pedersen current is induced in the $-x$ direction as the same way to the perturbed Hall current; see Figs.\ 2(f) and 3(c). Owing to effects of the new upward current at ($x$, $y$) = (53 km, 53 km), a turning point of the current flow pattern appears at $y \approx 50$ km.

\subsection{Comparison with previous studies}
The Cowling currents in our system are characterized by development of an eastward perturbed field $E_{1y}$ inside of the vortex. The field $E_{1y}$ is carried by Alfv$\acute{\rm e}$n waves from an external space to cancel the ambient ${\bm j}_{\rm H}(n_0, {\bm E}_0)$ and can be regarded as an east-west electric field tangential to the auroral arc proposed by Haerendel [2008]. And, when the tangential field exceeds a certain level, the vortex street forms leading to auroral breakup. Buchert and Budnik [1997] presented a similar scenario that an east-west induced electric field magnifies the upward currents. The relationship between the Pedersen and Hall currents inside of the vortex at the final stage $t/\tau_{\rm A}$ are essentially the same as a schematic view of classical Cowling channel patterns [Baumjohann, 1983; Yoshikawa et al., 2013]. That is, ${\bm j}_{\rm H}(n_1, {\bm E}_0) \parallel -y$, ${\bm j}_{\rm P}(n_0, {\bm E}_1) \parallel y$, and ${\bm j}_{\rm P}(n_1, {\bm E}_0)$ and ${\bm j}_{\rm H}(n_0, {\bm E}_1) \parallel -x$. Note that, on the other hand, the primary ${\bm j}_{\rm H}(n_1, {\bm E}_0)$ points poleward in the classical case originated from a tangential field. 

We could apply our current model to the situations in some previous studies. Fujii et al.\ [2012] presented the Cowling current system due to a secondary electric field related to a fully developed auroral arc. They supposed that there is an auroral arc accompanying an upward current and a downward current in the equatorward side, and these currents balance with the poleward Pedersen current ($\parallel {\bm E}_0$). They also required finiteness of the arc length and that the Pedersen and Hall currents flow in different altitudes. In these conditions, an eastward Hall current due to the primary field ${\bm j}_{\rm H}(n_1, {\bm E}_0)$ (hereafter, ${\bm j}_{\rm H10}$) is diverged. Since the divergence becomes much strong due to electron precipitation at the upward current region, new currents $j_\parallel < 0$ and $> 0$ are produced at the west/east edges of the arc, respectively. New currents flowing out to the Pedersen layer, an westward Pedersen current $j_{\rm P01}$, or an westward secondary electric field, is induced. They called this the Cowling current. 

Finiteness of the arc length may equal winding (vortices) of the arc-induced $j_\parallel > 0$ and $< 0$ in our calculation. From Figs.\ 2(c) and 3(f), we find that the current that balances to the pair currents is an eastward Hall current ${\bm j}_{\rm H10}$, and an westward Pedersen current ${\bm j}_{\rm P01}$ develops in its anti-parallel direction. This current system is consistent with that proposed by Fujii et al.\ [2012]. The new point of our system is that, inside of the vortex, there are an equatorward Hall current ${\bm j}_{\rm H01}$ induced by the pair currents $j_\parallel$ and an equatorward Pedersen current ${\bm j}_{\rm P10}$ due to polarization $n_1 < 0$. The above four currents are generally defined as a complete set of the Cowling current [Yoshikawa et al., 2013]. Because the arc length is not finite in our case, not so strong westward Pedersen current develops at the equatorward side arc at $x = 40$--45 km. 

Wild et al.\ [2000] presented a schematic interpretation of the horizontal currents related to auroral omega bands from observations of ion velocity by SuperDARN radar and ground-magnetometer. We expect that current closures have a similarity between omega bands in a scale of 500 km and vortex streets in a scale less than 50 km. They observed an enhancement of auroral luminosity (upward current) at the westward side of an omega band. They interpreted that the upward current balances with an westward Pedersen current (or electric field) and a downward current developed in the eastward side. The westward field produces a poleward $j_{\rm H}$ inside of the omega band, flowing around the upward/downward currents. They speculated that an ${\bm E} \times {\bm B}$ flow, anti-parallel to ${\bm j}_{\rm H10}$ in our terminology, corresponds to an observed poleward flow at a dark region between omega shapes. 

We argued their interpretation not to be unique. As shown in Fig.\ 3(f), the current that balances to the pair $j_\parallel$ on both sides of a vortex is not the Pedersen current but an westward Hall current $j_{{\rm H10}y} < 0$; it induces an eastward Pedersen current $j_{{\rm P01}y} > 0$ (or $E_{1y} > 0$). The Hall current ${\bm j}_{\rm H01}$ produced by $E_{1y} > 0$ points equatorward in our case, which is the opposite direction to the current ${\bm j}_{\rm H10}$ of Wild [2000]'s case. However, the field $E_{1y} > 0$ also produce a poleward flow at the westward side of the upward $j_\parallel$, which is consistent with the poleward flow in their observations. Our point is that a precise measurement of all variables ($n_1$, ${\bm E}_0$, ${\bm E}_1$, and $j_\parallel$) enables us to decide whether the currents around vortices are Pedersen-like or Hall-like.

\section{Conclusion}
As a series of studies from Hiraki [2015b], this study performed numerical simulations of Alfv$\acute{\rm e}$n waves destabilized in the MI coupling system and clarified the characteristics of the current system accompanied by auroral vortex streets; here, upward field-aligned current $j_\parallel$ of Alfv$\acute{\rm e}$n waves is considered to represent the auroral structure. The horizontal current is partitioned into the Pedersen and Hall currents and is further divided into two components due to the perturbed electric field ${\bm E}_1$ by Alfv$\acute{\rm e}$n waves and the perturbed electron density $n_1$ (or polarization). It was found that (i) at the initial stage, the arc-induced upward/downward $j_\parallel$ induce convergence/divergence of the polarized component of the Pedersen current; here, the ambient electric field ${\bm E}_0$ points poleward. (ii) A new upward $j_\parallel$ with a patchy structure is produced at equatorward of the downward $j_\parallel$ and balances with the polarized component of the Hall current. (iii) When a vortex street develops from the patch along with an winded downward $j_\parallel$, an upward $j_\parallel$ forms in the leading side of the vortex; it moves westward ($\parallel {\bm E}_0 \times {\bm B}_0$) in our setting. A downward $j_\parallel$ forming in the trailing side, these pair currents balance with an westward polarized component of the Hall current. It generates an eastward perturbed component of the Pedersen current inside of the vortex where there is no $j_\parallel$. On the other hand, counter-clockwise/clockwise patterns of the perturbed component of the Hall current form around the pair currents $j_\parallel > 0$ and $< 0$ (both edges of the vortex), respectively. The Hall current inside of the vortex points equatorward, and the polarized component of the Pedersen current is induced in the same direction. We expect that the complex current system of field-aligned and Pedersen/Hall currents yield a basic model to comprehend the realistic current systems accompanied by curl, spiral, and omega band structures observed at substorm periods.

\end{document}